\documentclass[onecolumn,pre,preprint,amsmath,amssymb,superscriptaddress]{revtex4-1}

\usepackage{amsmath}
\usepackage{bm}
\usepackage[dvipdfmx]{graphicx}
\usepackage{mathrsfs}
\usepackage{color}

\begin{document}

\title{Collision-induced torque mediates transition of chiral dynamic patterns formed by active particles}
\date{\today}

\author{Tetsuya Hiraiwa\footnote[1]{TH and RA equally contributed to this work. TH proposed the model and performed the numerical simulations, and RA interpreted the results and constructed the logic.}}
\email {mbithi@nus.edu.sg}
\address {Mechanobiology Institute, National University of Singapore, Singapore 117411, Singapore}
\address{Universal Biology Institute, The University of Tokyo, Hongo, Tokyo 113-0033, Japan.}
\author{Ryo Akiyama\footnotemark[1]}
\address {Department of Chemistry, Kyushu University, Fukuoka 819-0395, Japan}
\author{Daisuke Inoue}
\address{Faculty of Design, Kyushu University, Fukuoka 815-0032, Japan}
\author{Arif Md. Rashedul Kabir}
\address{Faculty of Science, Hokkaido University, Sapporo 060-0810, Japan}
\author{Akira Kakugo}
\address{Faculty of Science, Hokkaido University, Sapporo 060-0810, Japan}

\begin{abstract}
Controlling the patterns formed by self-propelled particles through dynamic self-organization is a challenging task. Although varieties of patterns associated with chiral self-propelled particles have been reported, essential factors that determine the morphology of the patterns have remained unclear. Here, we explore theoretically how a torque formed upon collision of the particles affects the dynamic self-organization of the particles and determine the patterns. Based on a particle-based model with a collision-induced torque and a torque associated with their self-propulsion, we find that introducing collision-induced torque turns the homogeneous bi-directional alignment of the particles into rotating mono-polar flocks, which helps resolve a discrepancy in the earlier observations in microfilament gliding assays.
\end{abstract}


\maketitle

\section{Introduction}
Controlling the patterns that emerge through dynamic self-organization of motile objects, as observed in the collective motion of living organisms \cite{vicsek2012collective,couzin2003, parrish2002}, has been a long-standing great challenge.
Recently, the dynamic patterns observed in nature have been contemplated in artificial systems by using the cytoskeletal filament microtubules (MTs) and F-actins. Upon propulsion by their associated biomolecular motors, the cytoskeletal filaments exhibited collective motion mediated by attractive interaction \cite{hess2005NL, tamura2011SM, kawamura2008chirality, kakugo2011controlled, wada2015sm, wada2015biomol} or crowded conditions \cite{schaller2010polar,inoue2015ns,sumino2012large}. Local alignment interactions, {\it i.e.}, collision-induced alignment of two self-propelled filaments, play important roles in the emergence of their collective motion  \cite{schaller2010polar,inoue2015ns,sumino2012large}. The resulting patterns of circular mesoscopic structures, streams, and vortices exhibited local or global rotational motion along the clockwise (CW) or counterclockwise (CCW) direction \cite{kawamura2008chirality,kakugo2011controlled,wada2015sm,wada2015biomol,sumino2012large, tanida2020gliding} that resembled the coherent motion exhibited by the living organisms in nature.

In nature, a rotational force or torque of cytoskeletal filaments
is known to play an important role in chiral morphogenesis of cells, tissues, and organisms \cite{naganathan2014active, tee2015cellular, novak2018NatCom}. Similarly, in the artificial systems made of cytoskeletal proteins, chiral collective behaviors have often been observed  \cite{kawamura2008chirality,kakugo2011controlled,wada2015biomol,sumino2012large,kim2018large,tanida2020gliding}.
Such observations and demonstrations motivated to explore the collective dynamics of chiral self-propelled objects theoretically by employing analytical and numerical approaches. Among these efforts, the most popular strategy was the introduction of chirality, {\it i.e.}, the left-right asymmetry, to the motion of the objects translocating in two dimensions \cite{Kuan2015PRErapid,Liebchen2017PRL,Chen2017Nat,Ai2018SM,Levis2018JOP,Lewis2019PRR,Zhang2021NP}.
Chirality was also introduced in relation to the mutual interactions of the objects \cite{Kruk2020PRE}.
A more straightforward
strategy has been to assume the objects with the finite-size and shape explicitly, instead of assuming the point-objects, and put chirality or LR asymmetry in the object shape \cite{Denk2016PRL,Liu2019SM,Moore2021SM}. 
Indeed, this approach has been useful in studying artificial systems in which the objects are of well-defined shape in a mechanistic sense; {\it e.g.}, gliding assay of cytoskeletal filaments \cite{Moore2021SM}. Such self-propelled objects with a chiral shape can exhibit varieties of dynamic self-organization.
In view of these two approaches to introduce chirality, the shape 
chirality can influence both the spontaneous motility and interaction of the objects. However, which chirality plays the crucial role in facilitating the emergence of
dynamic self-organization patterns has not been figured out yet. For example, recent study demonstrates well-defined chiral mono-polar flocking of MTs, or dense MT cluster in which motility directions of MTs are aligned unidirectionally, in a gliding assay on kinesins \cite{AfrozeInoue2021}, but it remains unclear how the chirality of MTs contributed to such mono-polar flocking. 
Indeed, the difference of chirality in spontaneous motility and interaction of the objects can be a hint of this as follows. Experimental observations in Ref. \cite{AfrozeInoue2021} suggested mono-polar flocking is attributed to chirality in interaction. In contrast, Refs. \cite{kim2018large,tanida2020gliding} report homogeneous bi-directional orientation of MTs with chiral rotational motion, in a gliding assay on kinesins, and such rotating bidirectional orientation can be explained by chirality in spontaneous motion \cite{tanida2020gliding}. Like this, to obtain a comprehensive understanding of the factors that  determine such differences in dynamic self-organization patterns with chirality, it appears inevitable to dissect the effect of chirality on the motility and interactions of self-propelled objects.

Here, we have demonstrated a 
systematic {\it in-silico} study on collective motion of self-propelled particles (SPPs) each of which has an intrinsic chirality of the both types as mentioned above; namely, chirality in self-propulsion and interaction. 
We consider SPPs  with intrinsic polarity where the SPPs move on a two-dimensional substrate. The SPPs interact with each other through isotropic core repulsion and bi-directional alignment. 
We chose such bi-directional alignment to avoid the emergence of mono-polar phase purely by the alignment interaction, which is the case for MTs in a gliding assay.
Two types of torque, self-propelled torque (ST) and collision-induced torque (CT)
(See Fig. \ref{graph:fig1} and below for more details), are applied to the particles as left-right (LR) asymmetric motility due to two-dimensionality.
We have investigated the emergence of patterns by tuning the strength of ST and CT without manipulating their alignment interactions. We found that 
when the CT is introduced, transition from 
bidirectional orientation 
to mono-polar flocking takes place although the alignment interaction is bi-directional. The emergence of mono-polar flocking mediated by chirality was reported in Ref. \cite{Moore2021SM}, where the authors investigated a mixture of two types of filaments having opposite chirality. 
On the contrary, the results presented in this article predict another mechanism to account for the chirality-induced mono-polar flocking, according to which such a combination of two types of filaments is unnecessary.

Before moving on, we recall the robust features experimentally observed in the collective motion of chiral MTs driven by kinesins, presented in Ref. \cite{AfrozeInoue2021}.
Chirality in the MT structure was introduced by polymerizing tubulins with a certain nucleotide, GMPCPP. 
The GMPCPP-MTs are found to align upon collision and eventually form mono-polar flocks. 
Notably, these flocks rotate dominantly in the CCW direction \cite{AfrozeInoue2021}.
In addition, when the MT-density is increased, mean curvature in a trajectory is also increased. 
These observations imply a correlation between the collision-induced torque and mono-polar flocking, which has motivated us to build up the simulation setting in this study.

\section{Model and method}

We consider $N$ SPPs, which are 
located at ${\bm x}_j =(x_j,y_j)$ ($j=1, 2, \cdots, N$)
and have intrinsic polarity ${\bm q}_j=(\cos \theta_j, \sin \theta_j)$ \cite{Hiraiwa14},
in a regular rectangle space with periodic boundaries in two dimensions \cite{Hiraiwa19,Hiraiwa20,Hayakawa20,tanida2020gliding}. 
We assume that 
locations and polarity directions of SPPs ($j=1, 2, \cdots, N$) obey
\begin{equation} \label{eq:vdyn-maintext}
{\bm \Theta} ({\bm q}_j) \frac{d {\bm x}_j}{dt} = v_0 {\bm q}_j + {\bm J^{\rm VE}}_j
\end{equation}
and 
\begin{align}
  \frac{d \theta_j}{dt} =
  {\bm J^{\rm AL}}_j \cdot {\bm q}_j^{\perp} + \xi_j
  + \omega^{\rm ST}  + {\Omega^{\rm CT}}_j \ ,
  \label{eq:thetadyn-maintext}
\end{align}
respectively, with ${\bm q}_j^{\perp}=(-\sin \theta_j, \cos \theta_j)$. 
Equation (\ref{eq:vdyn-maintext}) assumes the over-damped dynamics, and 
each object moves 
along the polarity ${\bm q}_j$ with a constant velocity $v_0$ \cite{Hiraiwa14}
in the absence of volume exclusion interactions.
The volume exclusion, by which two SPPs mechanically interact with each other, is implemented by
\begin{equation} \label{eq:volumeexclusion-maintext}
 {\bm J^{\rm VE}}_j = \beta \sum_{j' (n. j)} 
 \left( \frac{r}{|\Delta {\bm x}_{j,j'}|} - 1 \right) \frac{\Delta {\bm x}_{j,j'}}{|\Delta {\bm x}_{j,j'}|} \ . 
\end{equation}
The summation $\sum_{j' (n. j)}$ runs for all the neighbors $j'$ of $j$-th SPP, defined by
$|\Delta {\bm x}_{j,j'}|<r$ with $\Delta {\bm x}_{j,j'}= {\bm x}_j - {\bm x}_{j'}$.
Equation (\ref{eq:vdyn-maintext}) also assumes that
each SPP hardly moves along the direction perpendicular to the polarity direction,
using the rescaled anisotropic friction tensor ${\bm \Theta} ({\bm q}_j)= {\bm q}_j \otimes {\bm q}_j + R_{\zeta}^{-1} ({\bm I}  - {\bm q}_j \otimes {\bm q}_j)$ with the ratio $R_{\zeta}=\zeta_{\parallel}/\zeta_{\perp}$ of friction coefficients in parallel $\zeta_{\parallel}$ and perpendicular directions $\zeta_{\perp}$ \cite{tanida2020gliding}.
($\otimes$ is the tensor product, and ${\bm I}$ is the identity matrix.)
This anisotropy in friction has been implemented to phenomenologically reflect the observation that, when a gliding MT collides with another MT from its side, the colliding MT either stops to move or gets over the other, and the collided MT does not move into the direction perpendicular to its polarity \cite{tanida2020gliding}. This anisotropy is not essential for the phenomenology focused in this paper, as shown below in Appendix ("The case with isotropic mobility").
Equation (\ref{eq:thetadyn-maintext}) assumes
that polarity is spontaneously established for each SPP with a fixed amplitude, $|{\bm q}_i|=1$, and only its direction can evolve over time \cite{Hiraiwa14}.
The first term indicates bidirectional alignment interaction \cite{tanida2020gliding},
\begin{equation} \label{eq:nematicint-maintext}
 {\bm J^{\rm AL}}_j = 2 \alpha_{\rm AL} \sum_{j' (n. j)}  \left( {\bm q}_{j} \cdot {\bm q}_{j'} \right)  {\bm q}_{j'} \ .
\end{equation}
See Appendix ("Details of the theoretical model") for details.
The coefficient $\alpha_{\rm AL}$ indicates the strength of alignment.
The second term $\xi_j (t)$ represents Gaussian white noise with $\langle \xi_j \rangle =0$ and $\langle \xi_i(t) \xi_j (t') \rangle = 2 D \delta_{ij} \delta(t-t')
\label{eq:xidispersion}$
with the statistical average $\langle \cdot \rangle $.

\begin{figure}[t]
 \centering
 \includegraphics[width=12cm]{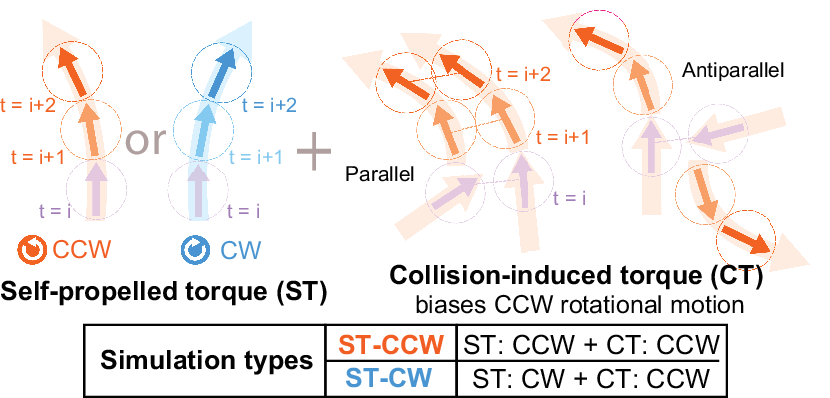}
 \caption{Schematic diagrams show the application of self-propelled 
 and collision-induced torque to SPPs.}
 \label{graph:fig1}
\end{figure}

In this study, we also apply two different types of torque, self-propelled torque (ST) and collision-induced torque (CT), to the particles,
which are represented by the last two terms in Eq.~(\ref{eq:thetadyn-maintext}) [Fig. \ref{graph:fig1}]. Note that, since SPPs are gliding on the substrate, the special directionality exists in the $z$-axis and here we are focusing on only the other two dimensions; reflecting this fact, we assumed torque as  representation of LR asymmetric motility. 
In ST, we assume that intrinsic polarity ${\bm q}_j$ of each SPP rotates with a given speed $\omega_{\rm ST}$ in either CCW and CW direction.
$\omega^{\rm ST}$ denotes the strength of ST [Fig.~\ref{graph:fig1} left], which we assume is a given constant. It is to be noted that ST has been observed in the
gliding assay of MTs \cite{kawamura2008chirality}. 
We further assume that, when SPPs collide, another torque is exerted on their intrinsic polarities, which we name CT. 
$\Omega^{\rm CT}_j$ denotes CT [Fig.~\ref{graph:fig1} right],
given by $\Omega^{\rm CT}_j=\omega^{CT} m_j$ with the number $m_j$ of SPPs within the range $r$ from the focused SPP and a constant $\omega^{CT}$ representing the strength of CT. Indeed, in the gliding assay of MTs on a kinesin coated surface in Ref \cite{AfrozeInoue2021},
an increase in the mean curvature of the trajectory was observed upon increasing the MT-density, which suggests that our assumption for CT is not artificial.
In the absence of these LR asymmetries, this model is essentially the same as that given in Ref. \cite{tanida2020gliding}.

We numerically calculate Eqs. (\ref{eq:vdyn-maintext})- (\ref{eq:nematicint-maintext}), after non-dimensionalization with characteristic length $X\equiv r$ and time $T \equiv r/v_0$. (For these purposes, we can simply put $r = 1$ and $v_0 = 1$.) We apply the Heun's method with a discretized time step $dt = 0.004$ up to variable total steps $M$. The value of $M$ or the corresponding time $t$ is mentioned in each figure legend. The parameters are set $R_{\zeta} =0.01$, $D=0.01$ (or P\'{e}clet number=$100$), $\alpha_{\rm AL}=1.0$ and $\beta=0.1$ unless otherwise mentioned. The number of objects and  the global object density are set to be $N=20,000$ and $\rho=0.2$, respectively, unless otherwise mentioned. The system size is $L=\sqrt{N/\rho}$ both for $x$ and $y$.

\section{Results}

\begin{figure}[!t]
 \centering
 \includegraphics[width=12cm]{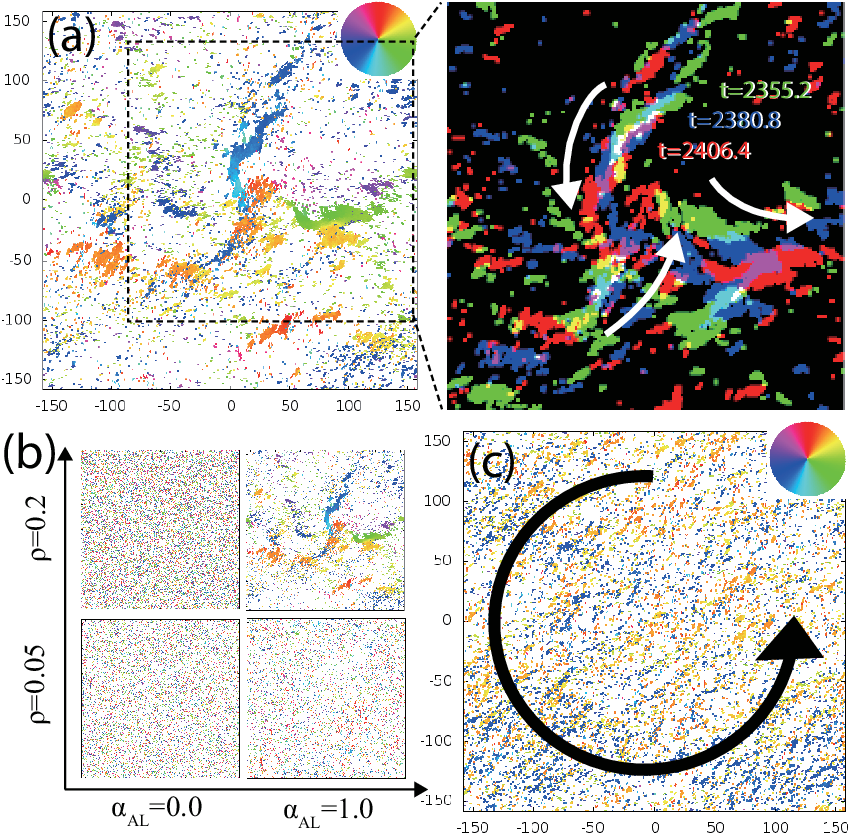}
 \caption{Typical dynamic patterns observed in the numerical simulations and {\it in vitro} gliding assay of MTs. See the text for details.
(a) Mono-polar flocking; for $\omega^{\rm CT}=0.001$ and $\omega^{\rm ST}=0.0$.
  Right: Time overlay images show movement of flocks with time. The green dots ($t=2355.2$) have moved to the blue dots ($t=2380.8$) and then to the red dots ($t=2406.4$).
  The white arrows show the traveling motions. Small structures have been removed by morphological erosion and dilation. 
(b) Snapshots of the numerical results for various SPP density $\rho$ and alignment strength $\alpha_{\rm AL}$. The other parameters are identical to (a). 
(c) Image shows rotating global bidirectional ordered state; for $\omega^{\rm CT}=0.0$ and $\omega^{\rm ST}=0.002$. $N=20,000$.
 Black arrow indicates the rotation direction of orientation.}
 \label{graph:fig2}
\end{figure}

First, we examine the possible patterns which may emerge based on our theoretical model system.
Typical snapshots are shown in Fig. \ref{graph:fig2}.
For a two-dimensional hard-disk system,
the formation of an ordered phase requires a packing fraction higher than 0.7 ($\rho=0.89$) 
\cite{Bernard11}. Despite a low global particle density, $\rho$ of 0.2,
mono-polar flocks
are observed {\it in silico}, as shown in Fig. \ref{graph:fig2}(a). The flocking is caused by the CCW-CT ($\omega^{CT}=0.001$, $\omega^{ST}=0.000$). 
The largest flock is found to be composed of several thousands of particles. 
Magnified time overlay images reveal the rotational motion of the flocks in the CCW direction [Fig. \ref{graph:fig2}(a)-right]. While rotating, the flocks collide with each other, and the particles scatter in all directions, which is followed by the regeneration of the flocks.
Note that when the density is low, the giant mono-polar flocks do not emerge [Fig. \ref{graph:fig2}(b)], which is consistent with the experimental results in Ref. \cite{AfrozeInoue2021}.
Simulating the case with only ST ($\omega^{CT}=0.000$, $\omega^{ST}=0.002$) exhibits homogeneous bi-directional orientation pattern which  always rotates counterclockwise, as shown in Fig. \ref{graph:fig2}(c),
consistent with those reported in previous literature \cite{kim2018large, tanida2020gliding}.

\begin{figure}[!t]
 \centering
 \includegraphics[width=15cm]{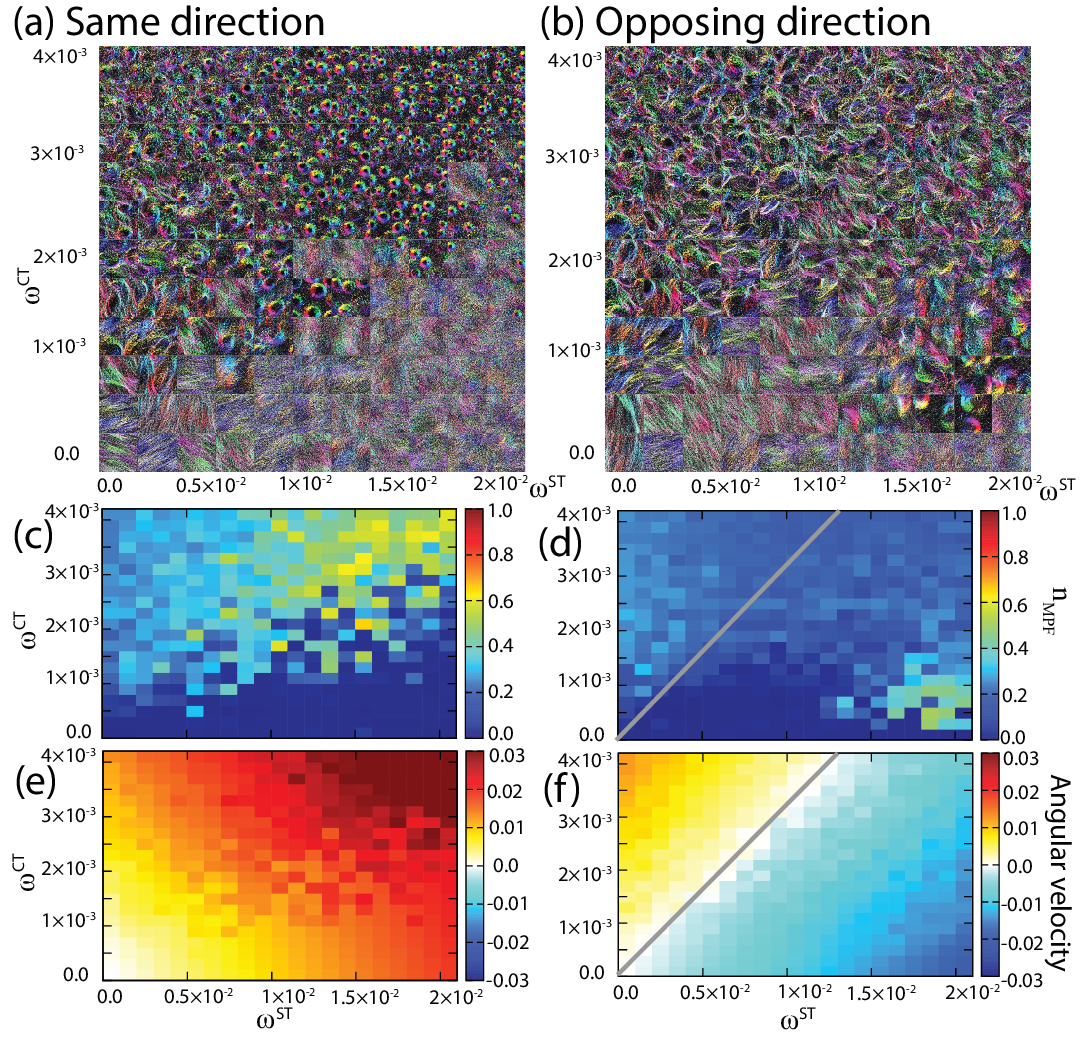}
  \caption{Dependence of dynamic patterns on the torque strengths $\omega^{\rm ST}$ and $\omega^{\rm CT}$. (a,b) Time-overlays. 
  (c,d)
  SPP number fraction within mono-polar flocks $n_{\rm MPF}$.
  To calculate, we first divide the system into $ 31 \times 31$ regions of interest (ROIs), and if a ROI has the polar order parameter $R \equiv | (\sum_{j \in \mathrm{ROI}} {\bm q}_j ) / (\sum_{j \in \mathrm{ROI}} 1) | $ higher than $R_{\rm th}=0.9$, 
  the ROI is regarded to be within mono-polar flocks.
  Giving the number of SPPs in all those ROIs $N_{\rm MPF}$, the color axis shows $n_{\rm MPF} = N_{\rm MPF}/N$.
  (e,f) Average angular velocity of each SPP.
  ST's direction is set to be (a,c,e) the same as and (b,d,f) the opposite of CT's direction (CCW).
  Gray lines
  highlight the parameters on which the angular velocity is zero. $M=1,280,000$.} 
\label{graph:fig3}
\end{figure}

To investigate the physical principle that controls the transition of these two distinct patterns,
we first examine how the two types of torque affect the pattern transition. 
The phase diagram against ST and CT in Fig. \ref{graph:fig3} (a, b) shows a correlation between morphology and the torques, and 
Figs. \ref{graph:fig3} (c, d) and (e, f) represent the number fraction of SPPs in mono-polar, and the angular velocities, respectively. 
The structures formed through the collective motion of SPPs 
are found to be dependent on both types of torque. 
From the time-overlay images, two distinct morphologies, mono-polar flocking phase and bi-polar phase, can be identified. The doughnut-like shapes, formed by the rotating flocks, represent the mono-polar phase. Some structures shown in fuzzy color, in which rotating flocks were not observed, represent stable bi-polar phase. For instance, rotating flocks are not observed when $\omega^{CT} = 0$. Since the effective attraction between particles is generated by the CT and alignment interaction,
these results seem reasonable. The bi-polar phase is observed for relatively high values of $\omega^{ST}$ and low values of $\omega^{CT}$ [Figs. \ref{graph:fig3}(a,b)].

When the ST and CT were opposite to each other, we found an island region in the lower right of Fig.~\ref{graph:fig3}(b) and (d) for the flocks rotating in the CW direction. The doughnut-like shapes in the island are the same as those in the mono-polar area on the upper left region in Fig. \ref{graph:fig3}(d). However, the SPP number fraction in the polar order region of the island area is larger than that of the upper left region. Thus, the collision probability in the flock increases, and the density increases by the collisions as $\omega^{ST}$ becomes large. 
 Therefore, the increase of the collision frequency caused by the ST leads to another mono-polar flocking phase in combination with the CT of the opposite direction.In contrast, when the CT is the same direction as the ST, as shown in Fig.~\ref{graph:fig3}(c), this island area is not observed. This is consistent with the above statement because the density seems to always decrease for increasing $\omega^{ST}$ in this case.

As mentioned above, the rotating flocks are observed for high values of $\omega^{CT}$ [Fig. \ref{graph:fig3}(a,b)]. As $\omega^{ST}$ increases, the boundary between the mono-polar flocking and homogeneous bi-polar phases moves upwards towards higher $\omega^{CT}$. This means ST suppresses the formation of rotating flocks, which is also supported by analysis of 
 the SPP number fraction in mono-polar flocks. 
 In case of the same direction [Fig. \ref{graph:fig3}(c)], the boundary appears for $\omega^{ST} < 2.0\times 10^{-2}$. In the case of opposing direction [Fig. \ref{graph:fig3}(d)], the boundary becomes unclear when $\omega^{ST} > 1.0\times 10^{-2}$ because of the existence of the exceptional region for the CW-rotating flocks.

 The boundary between the two phases appears as a thick band. 
 The map in the band is speckled,  
 which suggests that the mono-polar flocking and homogeneous phases are bi-stable (supported by hysteresis analysis in Appendix "Hysteresis analysis"). The speckled band is also found in the map of angular velocity for the case of the same direction [Fig. \ref{graph:fig3}(e)]. Similar to the fraction in mono-polar flocks, the angular velocity is relatively high in the mono-polar flocking phase [Fig. \ref{graph:fig3}(c)]. On the other hand, in Fig. \ref{graph:fig3}(f), the color bands are straight and parallel, which 
 indicates that the angular velocity is governed by the balance between $\omega^{CT}$ and $\omega^{ST}$. Thus, the angular velocity and the phase behavior are not strongly correlated when the ST and CT are opposite each other.

 In the case of opposing direction [Fig. \ref{graph:fig3}(f)], the $\omega^{CT}$ is positive (CCW) whereas the $\omega^{ST}$ is negative (CW). The white grids, shown by the 
 gray line,
 indicate that rotations are canceled by each other. 
 Furthermore, the slope of the line was roughly $0.3$.
We also evaluated the mean contact number around each particle in mono-polar flocks $\langle m_i \rangle_{i \in {\rm MPF} }$ in the numerical calculation, for the parameter window exhibiting the polar order, and it was around $2$ to $5$.
Hence, the above slope agrees with the condition for vanishing torque, $\omega^{ST}+\omega^{CT}\langle m_i \rangle_{i \in {\rm MPF} }\sim 0$. It is to note that, since the global particle density, $\rho$ has been set at $0.2$, this means the contact number and the slope are the consequence of flocking.

Finally, we discuss the mechanism behind the violation of bidirectional orientation and the formation of mono-polar flocks mediated by only the 
CT based on the three-SPP simulation (See Appendix "Three-particle simulation"
for the details). The increase of CT changes a stable state from bidirectional orientation to the rotating mono-polar flock [Fig.~\ref{graph:fig2}(a)].
Flocking is mediated by the alignment interaction and CT.
Although both 
the mono-polar and bidirectional 
orientation of motion can be stabilized by alignment interaction, simulations for a few SPPs in Fig. \ref{graph:3MTs}(a) and Appendix ("Three-particle simulation") reveal that, in the absence of CT, bi-polar orientation is stable and mono-polar flocks are rarely formed. 
On the contrary, CT rotates the direction of movement of the SPPs moving in the same direction as a cluster, which breaks the bi-polar orientation and provides more chances of mono-polar 
flocking [Fig.~\ref{graph:fig2}(b)].
For particles moving in the same direction, alignment interaction worked as an effective attractive interaction and  maintained the mono-polar flock once it is formed.
Therefore, when alignment interaction is strong enough and the density of SPPs is large enough, this effective alignment among the SPPs moving in the same direction may result in phase separation, and allows the emergence of the mono-polar flocks with high local density and large local polar order. In fact, such dependency on the associated parameters is seen in 
Appendix ("A few notes regarding density in flocks").
Note that we could not find the significant difference between the cases with and without ST [compare red circles and blue triangles in Fig.~\ref{graph:3MTs}(b)], 
which is a stark difference from the results shown in Fig.~\ref{graph:fig3}(c,d), where we simulated the many-SPP case ($N=20,000$) and observed the $\omega^{\rm ST}$-dependence of the SPP number fraction within mono-polar flocks. This difference implies that the many-body effect is indispensable to recapitulate the $\omega^{\rm ST}$-dependence of the threshold $\omega^{\rm CT}$ seen in Fig.~\ref{graph:fig3}.

\section{Conclusions}
In conclusion, through an {\it in silico} study, we have clarified
how the types of torque of SPPs, {\it i.e.} torque due to the collision and torque associated with their self-propulsion, can affect their coherent dynamics.
By varying the magnitudes of these two types of torque, we have discovered that there is a transition between different forms of coherent dynamics that are manifested by homogeneous bidirectional orientation and mono-polar flocking,
which have been observed in microtubule-kinesin gliding assays in Refs. \cite{kim2018large,tanida2020gliding} and Ref. \cite{AfrozeInoue2021}, respectively.
When the self-propelled torque is dominant, SPPs maintain their homogeneous bidirectional orientation, although 
their direction rotates.
We discovered that an increase in collision-induced torque breaks the homogeneous bidirectional order and the stabilized mono-polar flocks.
Our results clarify the role of collision-induced torque in the emergence of their coherent dynamics, 
and resolve the discrepancy in the observations mentioned above.
The findings in this work point out the importance of the type of torque as a control factor in the dynamic self-organization patterns of SPPs.

\section{Appendix}
\subsection*{Details of the theoretical model}

We consider $N$ particles in a square box with periodic boundaries in two dimensions,
and assume that each particle is a self-propelled particle (SPP)
with an intrinsic polarity along which the domain tries to move  \cite{tanida2020gliding}.
The location and intrinsic polarity of the $j$-th particle are described by ${\bm x}_j =(x_j,y_j)$ and ${\bm q}_j$, respectively ($j=1, 2, \cdots, N$).
We assume that the velocity ${\bm v}_j$, which determines the ${\bm x}_j$'s time evolution as
\begin{equation} \label{eq:veldef}
 \frac{d {\bm x}_j}{dt} = {\bm v}_j \ ,
\end{equation}
and the polarity ${\bm q}_j$ of the $j$-th particle obeys
\begin{equation} \label{eq:vdyn}
{\bm \Theta} ({\bm q}_j) {\bm v}_j = v_0 {\bm q}_j + {\bm J^{\rm VE}}_j
\end{equation}
and 
\begin{equation} \label{eq:qdyn}
 \frac{d {\bm q}_j}{dt} = {\bm J^{\rm AL}}_j + {\bm \xi}_j + {\bm \omega^{\rm ST}}  + {\bm \Omega^{\rm CT}}_j
\end{equation}
under the constraint $|{\bm q}_j|=1$, respectively, for every $j$.
Equation (\ref{eq:vdyn}) assumes the over-damped dynamics, and 
each particle moves with a constant velocity $v_0$ in the absence of volume exclusion interactions.
Equation (\ref{eq:vdyn}) also assumes that each particle hardly moves along the direction perpendicular to the direction of ${\bm q}_j$, which has been implemented by the (rescaled) anisotropic friction tensor ${\bm \Theta} ({\bm q}_j)= {\bm q}_j \otimes {\bm q}_j + R_{\zeta}^{-1} ({\bm I}  - {\bm q}_j \otimes {\bm q}_j)$ with the ratio $R_{\zeta}=\zeta_{\parallel}/\zeta_{\perp}$ of friction coefficients in parallel $\zeta_{\parallel}$ and perpendicular directions $\zeta_{\perp}$.
Here, $\otimes$ is the tensor product, and ${\bm I}$ is the identity matrix.
(The reason we introduced such anisotropy in friction is as follows: Ref. \cite{tanida2020gliding} observed that, when the gliding microtubule collides another microtubule from its side, the colliding microtubule either stops moving or gets over the other. In other words, phenomenologically, the microtubule collided from its side does not move into the direction perpendicular to its polarity. Since this work is motivated by the observations in microtubule gliding assay, as we detailed in Introduction, we reflected this fact phenomenologically by using the anisotropic mobility, or inverse friction, and below setting the friction perpendicular to the particle's intrinsic polarity to be much higher than the parallel counterpart, with which the particle hardly moves to the perpendicular direction indeed. However, it is to be noted that this anisotropic mobility is not essential in our main result.)
The term ${\bm \xi^{q}}_j(t)$ represents the noise, for which, for simplicity, we assume a white Gaussian noise with $\langle {\bm \xi^{q}}_j \rangle = 0$ and 
\begin{equation}
\langle \xi^{q}_{i,k}(t) \xi^{q}_{j,l}(t') \rangle = 2 D \delta_{ij} \delta_{kl} \delta(t-t') \ ,
\label{eq:xidispersion}
\end{equation}
where the subscripts $k$ and $l$ specify the directions, $k, l =x , y$,
with the statistical average $\langle \cdot \rangle $. 
The coefficient $D$ indicates the noise strength.
The particle-particle interactions are given by ${\bm J^{\rm VE}}_j$ and ${\bm J^{\rm AL}}_j$, which represent the volume exclusion and bidirectional alignment interaction with the interaction ranges $r$, respectively.
The volume exclusion is given by 
\begin{equation} \label{eq:volumeexclusion}
 {\bm J^{\rm VE}}_j = \beta \sum_{j' (n. j)} 
 \left( \frac{r}{|\Delta {\bm x}_{j,j'}|} - 1 \right) \frac{\Delta {\bm x}_{j,j'}}{|\Delta {\bm x}_{j,j'}|} \ . 
\end{equation}
The summation $\sum_{j' (n. j)}$ runs for all the neighbors $j'$ of $j$-th particles, defined by $|\Delta {\bm x}_{j,j'}|<r$ with $\Delta {\bm x}_{j,j'}= {\bm x}_j - {\bm x}_{j'}$. 
Here, we have assumed not rigorous but soft volume exclusion, and the softness is controlled by the factor $\beta$.
Bidirectional alignment interaction is given by \cite{tanida2020gliding} 
\begin{equation} \label{eq:nematicint}
 {\bm J^{\rm AL}}_j = 2 \alpha_{\rm AL} \sum_{j' (n. j)}  \left( {\bm q}_{j} \cdot {\bm q}_{j'} \right)  {\bm q}_{j'} \ .
\end{equation}
The coefficient $\alpha_{\rm AL}$ indicates the strength of alignment.
We can scrutinize the meaning of this term by rewriting it in the potential form; 
\begin{equation} \label{eq:JALvar}
 {\bm J^{\rm AL}}_j = - \frac{\partial V(\{{\bm q}_i\} )}{\partial {\bm q}_j}
\end{equation}
with
\begin{equation} \label{eq:JALpot}
V(\{{\bm q}_i\} = - \alpha_{\rm AL} \sum_{j,j'\in n.p.} | {\bm q}_j \cdot {\bm q}_{j'} |^2 = - \alpha_{\rm AL} \sum_{j,j'\in n.p.} \cos^2 (\theta_j - \theta_{j'})
\end{equation}
The summation $\sum_{j,j'\in n.p.}$ runs for all the neighboring $j$-th and $j'$-th particle pairs, defined again by $|\Delta {\bm x}_{j,j'}|<r$. As Eqs. (\ref{eq:JALvar}) and (\ref{eq:JALpot}) indicate, this interaction term align the polarities of neighboring pair of particles in the bidirectional way, {\it i.e.} toward either $\theta_j - \theta_{j'}=0$ or $\theta_j - \theta_{j'}=\pi$. 
In this study, we naively assume given constants $\alpha_{\rm AL}$ and $\beta$ for the interaction, but there are other possible choices. For example, in some literature on collective motion of SPPs,
the interaction is defined in the way that its strength depends on the SPP speed \cite{Baskaran18}. This may be the point which one has to be careful when investigating the phase diagram over {\it e.g.} $v_0$. In this paper, we focus on only the torque strengths, mentioned below, so that this choice may not affect the conclusion.

The last two terms of Eq. (\ref{eq:qdyn}) are the key terms of this study, which give rise to chirality. As mentioned in Introduction, in this article, we assume that chirality 
affects the system's dynamics through
the two different ways: one is the self-propelled torque (ST)
${\bm \omega^{\rm ST}} \equiv \omega^{\rm ST} {\bm q}_j^{\perp} $, and the other is the collision-induced torque (CT)
${\bm \Omega^{\rm CT}}_j \equiv \Omega^{\rm CT}_j {\bm q}_j^{\perp}$. 
Here, ${\bm q}_j^{\perp}$ is one of the unit vectors perpendicular to the polarity, given by ${\bm q}_j^{\perp} \equiv (- \sin \theta_j, \cos \theta_j)$. 
The constant $\omega^{\rm ST}$ is the strength of ST.
On the other hand, 
$\Omega^{\rm CT}_j$ is given by $\Omega^{\rm CT}_j=\omega^{CT} m_j$ with the number $m_j$ of particles within the range $r$ from the focused particle ($j$), and the constant $\omega^{CT}$ controls the strength of CT.
(See Fig. \ref{graph:fig1} for the schematics.)
It is worth noted that, in two dimensions, there are two unit vectors perpendicular to a certain reference vector, corresponding to left or right. The above definition of ${\bm q}_j^{\perp}$ selected out one of them. Thus, the existence of ${\bm q}_j^{\perp}$ in the definitions of ${\bm \omega^{\rm ST}}$ and ${\bm \Omega^{\rm CT}}$ is expressing the origin of chirality.

Equation (\ref{eq:qdyn}) can be rewritten as the time evolution of the angles $\theta_j$ of the polarity directions, using ${\bm q}_j = (\cos \theta_j, \sin \theta_j)$.
For this purpose, we may take the inner products of ${\bm q}_j^{\perp}$ and the both sides of Eq. (\ref{eq:qdyn}). 
As a result, Eq.~\ref{eq:thetadyn-maintext} is obtained, with $\xi_j \equiv {\bm \xi}_j \cdot {\bm q}_j^{\perp} $. 
The new noise term $\xi_j(t)$ represents the angular noise, and since $\xi_j = - \xi^{q}_{x,j} \sin \theta_j + \xi^{q}_{y,j} \cos \theta_j$,
it is again a Gaussian white noise with $\langle \xi_j \rangle =0$ and
\begin{equation} \label{eq:thetaxidispersion}
\langle \xi_{\alpha}(t) \xi_{\alpha'} (t') \rangle = 2 D \delta_{{\alpha},{\alpha}'} \delta (t-t') .
\end{equation}
Note that $D^{-1}$ now characterizes the persistence time of polarity direction, or characteristic time of angular fluctuation.

Equation (\ref{eq:vdyn}) and Eq. \ref{eq:thetadyn-maintext}, {\it i.e.}, the angular description of Eq. (\ref{eq:qdyn}), can be rewritten by the dimensionless forms as
\begin{equation} \label{eq:vdyn-dimles}
 \frac{d \tilde{\bm x}_j}{d \tilde{t}} = 
 {\bm q}
 + \tilde{\beta} \sum_{j' \ (n. j)} \frac{ \tilde{\Delta \bm x}_{j,j'} }{ | \tilde{\Delta \bm x}_{j,j'}|^2 } 
\end{equation}
and 
\begin{equation} \label{eq:polaritydynamics-dimles}
 \frac{d \theta_j}{d \tilde{t}} =  2 \tilde{\alpha}_{\rm AL} \sum_{j' (n. j)}  \left( {\bm q}_{j} \cdot {\bm q}_{j'} \right) ({\bm q}_j^{\perp} \cdot {\bm q}_{j'})
 + \tilde{\xi}_j + 
   \tilde{\omega^{\rm ST} } + \tilde{ {\Omega^{\rm CT}}_j }
   \ ,
\end{equation}
respectively, 
where $\tilde{\bm x}_j = {\bm x}_j/X$, $\tilde{t}=t/T$, $\tilde{\Delta \bm x}_{j,j'} = \Delta {\bm x}_{j,j'}/X$, $\tilde{\beta} = \beta T/X$, $\tilde{\alpha}_{\rm AL} = \alpha_{\rm AL} T$,
$\tilde{\omega^{\rm ST} }=\omega^{\rm ST} T$
and $\tilde{ {\Omega^{\rm CT}}_j }= {\Omega^{\rm CT}}_j T$,
with characteristic length $X\equiv r$ and time $T \equiv r/v_0$.
The noise term is also rescaled into the new notation $\tilde{\xi}_j(t)$, which is a Gaussian white noise satisfying $\langle \tilde{\xi}_j \rangle =0$ and
\begin{equation} \label{eq:thetaxidispersion}
\langle \tilde{\xi}_{\alpha}(t) \tilde{\xi}_{\alpha'} (t') \rangle = 2 \tilde{D} \delta_{{\alpha},{\alpha}'} \delta (\tilde{t}-\tilde{t'}) .
\end{equation}
with $\tilde{D}=DT$.
In the main text, we applied the same nondimensionalization by putting $r = 1$ and $v_0 = 1$.
The propulsion strength is often quantified by using P\'{e}clet number
\cite{Redner13,Kuan2015PRErapid}.
The P\'{e}clet number, or specifically the rotational P\'{e}clet number \cite{Kuan2015PRErapid}, is defined by ${\rm Pe} \equiv v_0 \tau_p /r$, with the migration persistence time $\tau_p$ \cite{Redner13}.
${\rm Pe}$ is given by the inverse of the dimensionless noise, ${\rm Pe} = \tilde{D}^{-1}$, because $\tau_p = D^{-1}$ in our case.
The parameter values which we used in our simulations here correspond to ${\rm Pe} = 100$.

\subsection*{Hysteresis Analysis}
In Fig.~\ref{graph:fig4}(a), the SPP number fraction in mono-polar flocks, $n_{\rm MPF}$, is plotted with changing $\omega^{\rm CT}$ up and down, which shows the clear hysteresis. This result suggests that the emergence mono-polar flocking from the homogeneous bi-directional orientation (or {\it vice versa}) 
has the first order transition-like nature. This is consistent with the appearance of thick transition region in Fig. \ref{graph:fig3}.

\subsection*{A few notes regarding density in flocks}
In the case that ST and CT are the same directions with each other (filled circles) in  Fig. \ref{graph:fig4}(b), the local density in mono-polar flocks is between $0.4$ to $1.5$ with the median of $~1.0$. When the density is $1.0$, the packing is almost the closest. The density increases as the absolute value of the average angular velocity increases. This result suggests that the collisions generate the effective attraction and mono-polar flocks are maintained by the collisions.

The density increases as the absolute value of the mean rotation speed increases. The collision becomes more frequent and stronger as the torque becomes greater. We can regard that the effective attraction between the particles is strong under such conditions. In the case that ST and CT are the opposite directions with each other (filled diamonds in Fig. \ref{graph:fig4}(b)), this feature is remarkable. Therefore, we find the density $2.5$ or more. The large value suggests, the particles overlay each other due to the strong collision. Here, the red diamonds for the high-density flocks belong to the island region of monopolar flocks in Fig.~\ref{graph:fig3}(b) and (d) because the red diamonds mean $\omega^{ST} > 1.5 \times 10^{-2}$. The strong collisions in the cluster caused by the CW-ST and the CCW-CT are confirmed in Fig. \ref{graph:fig4}(b). These high-density flocks are clearly distinguished from the low-density homogeneous phase.

\subsection*{Three-particle simulation}
Figure~\ref{graph:3MTs-App} provides additional results of numerical simulations for dynamics of three particles in a regular square with periodic boundaries. 
(The system width is set to be $L=\sqrt{N/\rho}$ with $N=3$ and $\rho=0.02$.)
In the simulations here, the initial locations and polarities of the three SPPs are set in a triangular and inwardly-pointing manner, respectively, which allows three SPPs to effectively collide, as shown in Fig.~\ref{graph:3MTs-App}(a) top left; $t=0$. 
Typical snapshots of the simulation results are shown in Fig.~\ref{graph:3MTs-App}(a,b).
Here, we have applied $\omega^{\rm CT}=0.1$, which is much larger than the maximum strength used in the main text, $\omega^{\rm CT}=0.004$.
As shown in Fig.~\ref{graph:3MTs-App}(a), the three SPPs show various pair/triplet dynamics including the bi-directional orientation, merging into a single mono-polar flock, and split of the flock.
In the main text, we found that the collision-induced torque can induce the mono-polar flocking.
Figure~\ref{graph:3MTs-App}(b) indeed demonstrates the case when a two-particle cluster rotates, which broke the bidirectional orientation, and results in formation of the mono-polar flock of three SPPs.

\begin{figure}[!t]
 \centering
 \includegraphics[width=12cm]{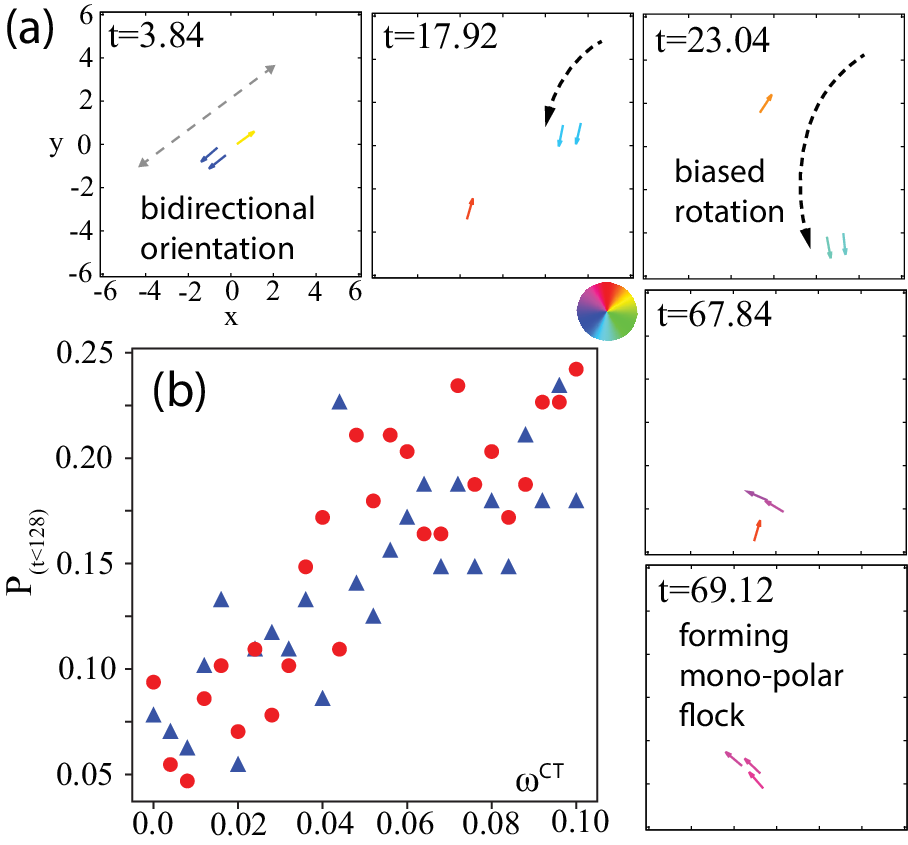}
 \caption{ Numerical simulations for dynamics of three particles in a regular square with periodic boundaries. (a) Snapshots showing typical time-evolution. Each colored arrow represents the location and polarity direction of each element,
and the color indicates its polarity direction corresponding to the color wheel. 
We here apply $\omega^{\rm CT}=0.1$, which is much larger than the maximum strength used in the main text, $\omega^{\rm CT}=0.004$. 
$\omega^{\rm ST}=0.0$, and the other parameters are the same as in the main text. A two-particle cluster rotated, which broke the bidirectional orientation, and resulted in the formation of the mono-polar flock. 
(b) $\omega^{\rm CT}$-dependence of the probability by which the three SPPs form the mono-polar flock at least a time by $t=128$.
Red circles: $\omega^{\rm ST}=0.0$. Blue triangles: $\omega^{\rm ST}=0.02$.}
\label{graph:3MTs}
\end{figure}

\begin{figure}[!t]
 \centering
 \includegraphics[width=14cm]{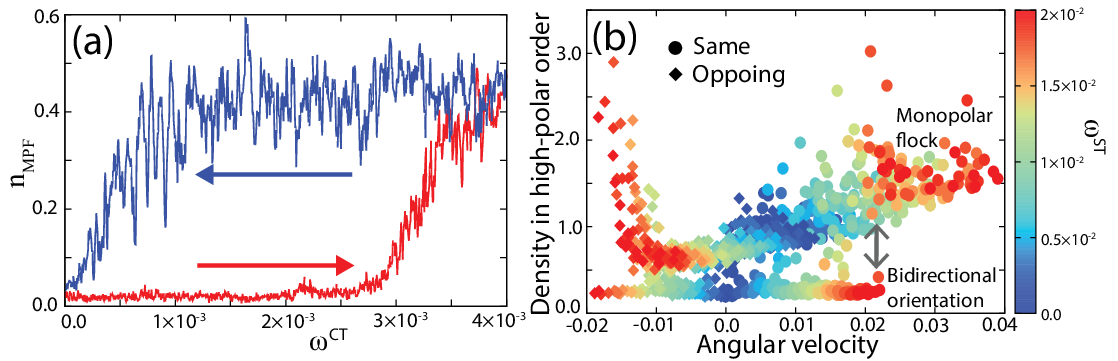}
  \caption{First order transition-like nature of
  mono-polar flocking. (a) Hysteresis of SPP number fraction in 
  mono-polar flocks, $n_{\rm MPF}$. $\omega^{\rm ST}$ is set to be $0.01$, and $\omega^{\rm CT}$ was swept from $0$ to $0.004$ (ascent; red curve) and, after that, {\it vice versa} ( descent; blue curve). See the legend of 
  Fig. \ref{graph:fig3}(c,d) 
  for the definition of $n_{\rm MPF}$.
  (b) Local density in 
  regions with high polar order, against the average angular velocity of each SPP. The color of each mark indicates the CT strength $\omega^{\rm ST}$ whereas the shape indicates the direction of ST (circles and diamonds; the same as and opposite to CT, respectively). Multiple marks with the same color and shape correspond to various $\omega^{\rm CT}$.
} 
\label{graph:fig4}
\end{figure}

\begin{figure}[!h]
 \centering
 \includegraphics[width=11cm]{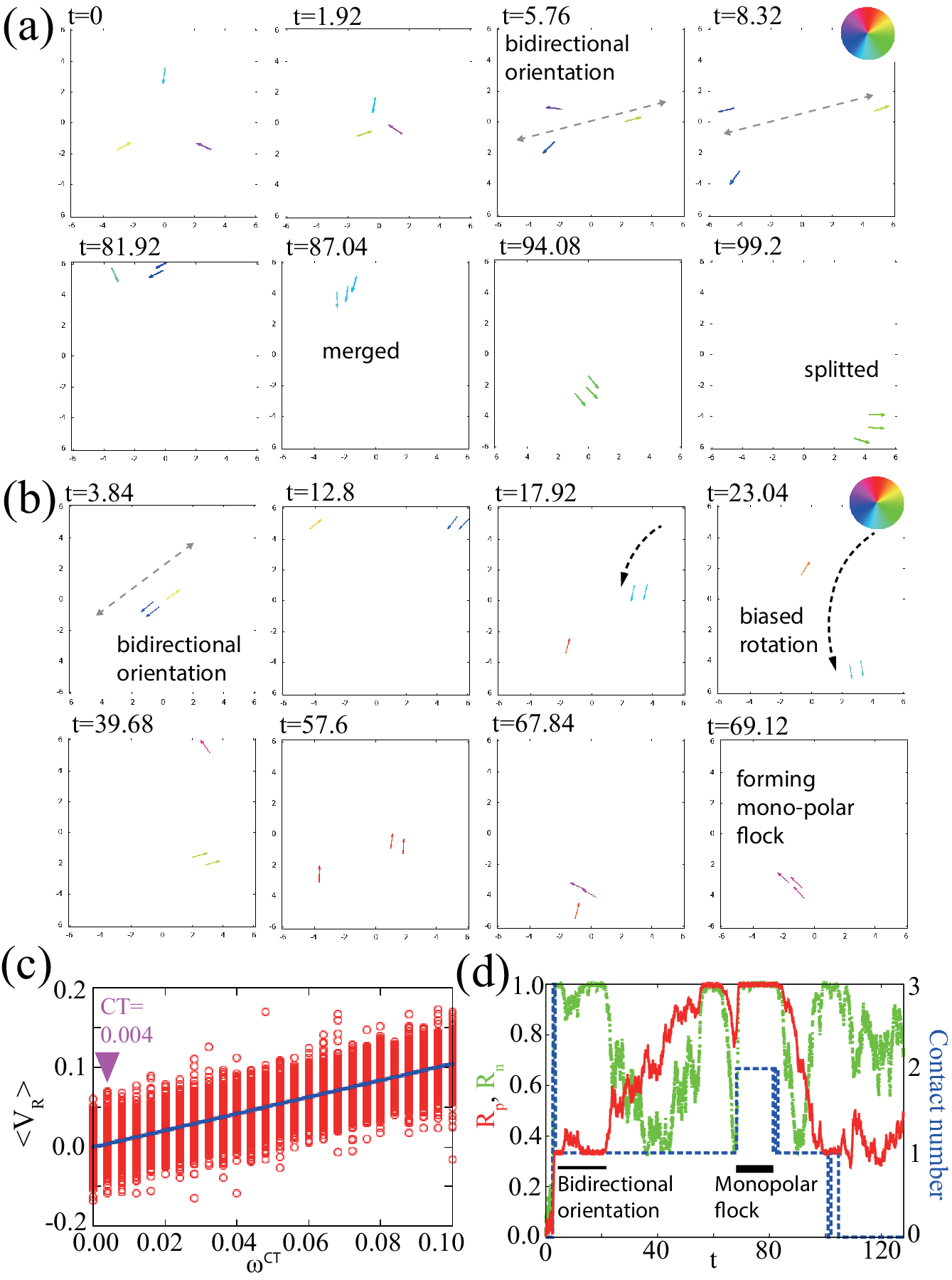}
 \caption{Numerical simulations for dynamics of three particles in a regular square with periodic boundaries.
 (a) Snapshots for a single simulation. Each colored arrow represents the location and polarity direction of each particle,
and the color indicates its polarity direction corresponding to the color wheel. 
$\omega^{\rm CT}=0.1$, $\omega^{\rm ST}=0.0$, and the other parameters are the same as in the main text.
 (b) Another sample with the same parameter values with (a).
 (c) Rotation velocity $V_R$ of intrinsic polarities during the particle-particle contact, averaged over all particles and time $\langle \cdot \rangle$ vs $\omega^{\rm CT}$.
 Different marks corresponding to different runs.
  The plots are fitted by $\langle V_R \rangle = (1.048 \pm 0.002) \omega^{\rm CT} + (-0.00016 \pm 0.00011)$.
  $2,048$ runs were simulated in total for each $\omega^{\rm CT}$, and each simulation was carried out up to $t=256$.
(d) Time evolution of polar order $R_p$ (red solid curve), nematic order $R_n$ (green dotted curve) and contact numbers (blue broken curve) for the simulation sample 
identical to (b).
}
\label{graph:3MTs-App}
\end{figure}

The results of such three-SPP simulations are quantified in Figs.~\ref{graph:3MTs}(b) and \ref{graph:3MTs-App}(c,d).
Figure~\ref{graph:3MTs-App}(c) plots rotation velocity $V_R$ of intrinsic polarities during the particle-particle contact, averaged over all elements and time ($\langle \cdot \rangle$), against $\omega^{\rm CT}$.
It indeed increases linearly for increasing $\omega^{\rm CT}$ on average. 
In Fig.~\ref{graph:3MTs-App}(d), 
time evolution of the polar order $R_p$ (red solid curve), nematic order $R_n$ (green dotted curve) and contact numbers (blue broken curve)  
corresponding to the sample dynamics shown in Fig.~\ref{graph:3MTs-App}(b), or Fig.~\ref{graph:3MTs}(a).
The polar- and nematic-order parameters, $R_p$ and $R_n$, are defined as
\begin{equation}
    R_p(t) \equiv \left | \sum_{i=1,2,3} \exp \left[ i \theta_i (t) \right] \right | = \left | \sum_{i=1,2,3} {\bm q}_i (t) \right |
\end{equation}
and
\begin{equation}
    R_n(t) \equiv \left | \sum_{i=1,2,3} \exp \left[ 2 i \theta_i (t) \right] \right | = \left | \sum_{i=1,2,3} \left[ q_x^2(t) - q_y^2(t) + 2i q_x(t) q_y(t) \right] \right |
\end{equation}
respectively. The summation $\sum_{i=1,2,3}$ runs over three SPPs labelled by $i=1,2,3$. 
We assume that two SPPs are in contact with each other if the distance of those two SPPs is smaller than the interaction range $r$, 
and contact number is defined as the number of such pairs in contact.
Figure~\ref{graph:3MTs-App}(d) also shows the mono-polar flock and bidirectional orientation regimes.
Here, the mono-polar flock regime has been defined
as the time window during which $R_p > R_{\rm Th}$ and the contact number is $2$ or $3$.
The bidirectional orientation regime has been defined as the time window during which $R_n > R_{\rm Th}^2$, $R_p < R_{\rm Th}$ and the contact number is equal to or higher than $1$.
(We here set $R_{\rm Th}=0.9$ again.)
Although there is another short bidirectional orientation regime around $t=100$ in Fig.~\ref{graph:3MTs-App}(d), we skipped plotting it for better visibility. 
In Fig.~\ref{graph:3MTs}(b), 
we plotted the probability $P_{(t<128)}$ by which the three SPPs form the mono-polar flock at least a time by $t=128$ for various $\omega^{\rm CT}$ and $\omega^{\rm ST}$.
To define $P_{(t<128)}$ for each parameter set, we have counted the number of the samples which has mono-polar flock regimes at least a time until $t=128$
and divided it by the total number of samples ($128$ samples).

\subsection*{The case with isotropic mobility}

\begin{figure}[!t]
 \centering
 \includegraphics[width=14cm]{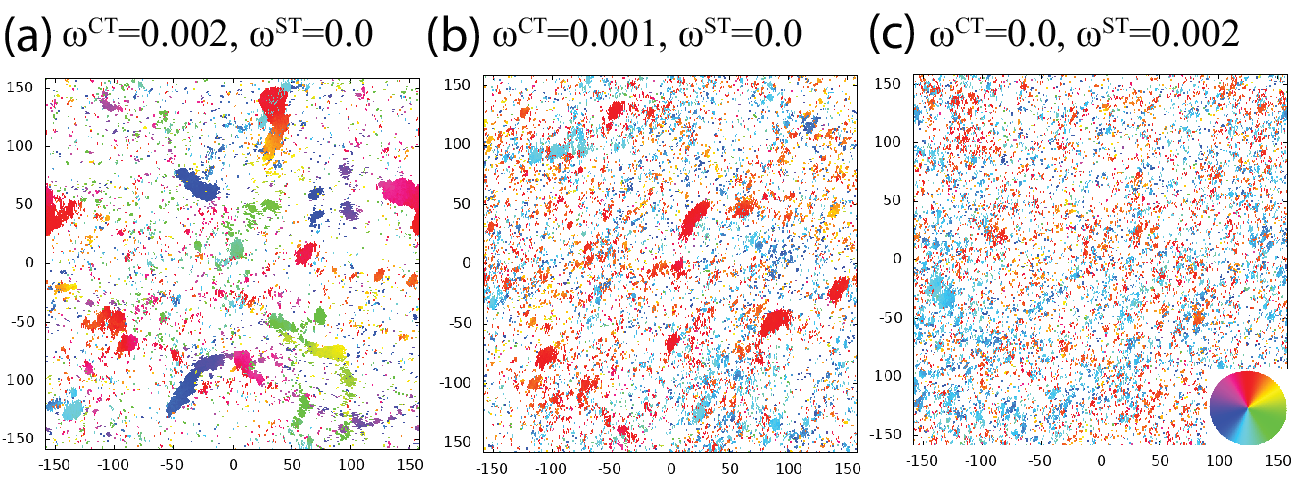}
  \caption{Dynamic patterns for the case with isotropic friction, $R_{\zeta}=1.0$. Typical snapshots of the numerical results are shown for (a) $\omega^{\rm CT}=0.002, \omega^{\rm ST}=0.0$, (b) $\omega^{\rm CT}=0.001, \omega^{\rm ST}=0.0$, and (c) $\omega^{\rm CT}=0.0, \omega^{\rm ST}=0.002$. $\rho=0.2$, $\alpha_{\rm AL}=1.0$, and $N=20,000$. Except that the friction is isotropic, these settings are corresponding to those in Fig.~\ref{graph:fig2}. The mono-polar flocking and bidirectional orientation for the cases with only the CT and only the ST, respectively, are reproduced while typical morphology of each mono-polar flock seems different from that seen in the anisotropic-friction case, where the flock seems to be more elongated (Fig.~\ref{graph:fig2}).}
\label{graph:isotropic}
\end{figure}

As expected by the explained mechanism, the formation of mono-polar flocks by CT is not relying on the anisotropic setting of friction in Eq.~(\ref{eq:vdyn-maintext}) or (\ref{eq:vdyn}). In Fig.~\ref{graph:isotropic}, we performed the simulations with the isotropic setting, $R_{\zeta}=1.0$, and indeed obtained the similar snapshots.


\section*{Acknowledgements}
We thank S. Tanida and M. Sano for valuable discussions in the TH's previous works which help us design this work.
We also thank Yuting Lou, Rakesh Das, Alok Ghosh and Ayumi Ozawa for helpful comments on this work.
This work was supported by the Mechanobiology Institute, National University of Singapore, (to TH), the JSPS KAKENHI grant number JP16K17777, JP19K03764 (to TH), JP19H01863, JP19K03772, JP18K03555, JP16K05512 (to RA), JP20K15141, JP21H05886 (to DI), JP20H05972, JP21K04846 (to AMRK), and JP18H03673 (to AK), a Grant-in-Aid for Scientific Research on Innovative Areas "Molecular Engine" (JSPS KAKENHI Grant Number JP18H05423) and a Grant-in-Aid for JSPS Research Fellows  (JP18F18323) (to AK), "Leading Initiative for Excellent Young Researchers (LEADER)" (JSPS Grant number RAHJ290002) (to DI), a research grant from Hirose Foundation  (PK22201017) (to AMRK), and New Energy and Industrial Technology Development Organization (NEDO) (JPNP20006) (to AK).

\section*{Conflicts of interest}
There are no conflicts to declare.

\bibliography{Reference.bib}

\clearpage

\clearpage

\clearpage

\clearpage

\clearpage

\clearpage

\clearpage

\end{document}